\documentclass[12pt]{article}  
\usepackage{amsmath}
\usepackage{amsfonts,amssymb}

\newcommand{\cH}{{\mathcal H}}

\newcommand{\cF}{{\mathcal F}}

\newcommand{\cL}{{\mathcal L}}

\def\R{{\mathbb R}}
\def\N{{\mathbb N}}

\newcommand{\Arp} {A_{\rho, +}}
\newcommand{\Arm} {A_{\rho, -}}
\newcommand{\Arpm} {A_{\rho, \pm}}
\newcommand{\Brp} {B_{\rho, +}}
\newcommand{\Brm} {B_{\rho, -}}
\newcommand{\Brpm} {B_{\rho, \pm}}
\newcommand{\harp} {\hat{A}_{\rho, +}}
\newcommand{\harm} {\hat{A}_{\rho, -}}
\newcommand{\harpm} {\hat{A}_{\rho, \pm}}
\newcommand{\hbrp} {\hat{B}_{\rho, +}}
\newcommand{\hbrm} {\hat{B}_{\rho, -}}
\newcommand{\hbrpm} {\hat{B}_{\rho, \pm}}
\newcommand{\rhoe}{\rho^{1/2}}

\newtheorem{theorem}{Theorem}[section]

\newtheorem{proposition}{Proposition}[section]
\newtheorem{lemma}{Lemma}[section]

\newtheorem{assumption}{Assumption}[section]
\newtheorem{remark}{Remark}[section]

\newenvironment{proof}{\medskip\par\noindent{\it Proof:}}{$\square$\newline\smallskip}

\makeatletter
    
    \@addtoreset{equation}{section}
\makeatother

\makeatother 
\begin{document}

\title{{Improvement of Uncertainty Relations for Mixed States}}

\author{Yong Moon Park}
\date{
  { \small Department of Mathematics, 
  \\
  Yonsei University, Seoul 120-749, Korea \\
 E-mail : ympark@yonsei.ac.kr }}



\maketitle
\begin{abstract}
We study a possible improvement of uncertainty relations. The
Heisenberg uncertainty relation employs commutator of a pair of
conjugate observables to set the limit of quantum measurement of
the observables. The Schr\"odinger uncertainty relation improves
the Heisenberg uncertainty relation by adding the correlation in
terms of anti-commutator. However both relations are insensitive
whether the state used is pure or mixed. We improve the
uncertainty relations by introducing additional terms which
measure the  mixtureness of the state. For the momentum and
position operators as conjugate observables and for the thermal
state of quantum harmonic oscillator, it turns out that the
equalities in the improved uncertainty relations hold.
\end{abstract}

\baselineskip=18pt
\section{Introduction}
Soon after Heisenberg and Schr\"odinger invented Quantum mechanics
around 1925, Heisenberg discovered the uncertainty relation in
1927 \cite{Hei}. The standard form of Heisenberg's uncertainty
relation for any pair of obsevables $A$ and $B$ and a density
matrix $\rho$ is expressed as
\begin{equation}\label{1.1}
\frac14 | \langle [A, B] \rangle _\rho |^2 \le Var_\rho (A)
Var_\rho (B),
\end{equation}
where $Var_\rho(A) =tr(\rho A^2) - (tr (\rho A))^2$ is the
variance of $A$ in the state defined by $\rho$, and $Var_\rho (B)$
is defined analogously, $\langle [A, B ]\rangle _\rho = tr (\rho
[A,B])$ is the expectation of the commutator $[A,B] =AB-BA$. The
relation (\ref{1.1}) states the fundamental limitation on quantum
measurement for incompatible (non-commuting) observables and has
been played fundamental role in quantum theory.

In 1930, Schr\"odinger improved the uncertainty relation
(\ref{1.1}) by including the correlation between observables :
\begin{equation}\label{1.2}
\frac14 | \langle [A, B] \rangle_\rho |^2 + \frac 14 | \langle \{
A_0, B_0 \} \rangle_\rho |^2 \le Var_\rho (A) Var_\rho(B),
\end{equation}
where $\langle \{ A_0, B_0 \} \rangle _\rho = tr(\rho \{ A_0, B_0
\} )$ denotes the expectation of anti-commutator $\{ A_0, B_0 \}
=A_0 B_0 + B_0 A_0$ and $A_0 =A- \langle A\rangle_\rho$, $B_0=B -
\langle B\rangle_\rho$. The first term in the left hand side of
(\ref{1.2}) encodes incompatibility, while the second term encodes
correlation between observables $A$ and $B$.

In recent years in the field of quantum computation and quantum
information, the strong correlation, such as the phenomenon of
entanglement, in the quantum world that can not be occured in
classical mechanics, has intensively studied \cite{NC}. Thus one
expects that the Schr\"oinger uncertainty relation will be played
an important role in quantum theory \cite{Suk}.

In this paper, we improved the uncertainty relations (\ref{1.1})
and (\ref{1.2}) by introducing additional terms in the lower
bounds of (\ref{1.1}) and (\ref{1.2}) respectively. We will show
that for any observables $A$ and $B$, and any density matrix
$\rho$, the following uncertainty relations hold:
\begin{eqnarray} \label{1.3}
\begin{aligned}
\frac 14 | \langle [A, B]\rangle_\rho |^2 +  & tr (A_0 \rho^{1/2}
A_0
\rho^{1/2}) ( tr (B_0  \rho^{1/2} B_0 \rho^{1/2})) \\
&  \le Var_\rho (A) Var_\rho (B)
\end{aligned}
\end{eqnarray}
and
\begin{eqnarray} \label{1.4}
\begin{aligned}
\frac 14 | \langle [A, B]\rangle_\rho\,|^2 + &\frac 14 | \langle
\{A_0,
B_0 \} \rangle_\rho\,|^2  + M(A_0, B_0 ; \rho)  \\
& \le  Var_\rho (A) Var_\rho (B),
\end{aligned}
\end{eqnarray}
where the quantity $M(A_0, B_0 ; \rho)$ is defined in Theorem
\ref{theorem2.2} explicitly. Notice that the relation (\ref{1.3})
and (\ref{1.4}) are improved version of the relation (\ref{1.1})
and (\ref{1.2}) respectively. If the density matrix $\rho$ is
pure, the the second term in the left hand side of (\ref{1.3}) and
the third term in (\ref{1.4}) are vanished and so (\ref{1.3}) and
(\ref{1.4}) are reduced to the original relations (\ref{1.1}) and
(\ref{1.2}) respectively.

It may be worth to mention that for any observable $A$ the
functional
$$
\rho \mapsto tr (A\rho^{1/4} A \rho^{1/4})
$$
is concave by Lieb's concavity theorem \cite{Lie}, and so in a
sense the values of additional terms in the above measure the
mixtureness of $\rho$. We also note that the Wigner-Yanase skew
information \cite{WY} is given by
\begin{eqnarray} \label{1.5}
\begin{aligned}
I(\rho, A) &= -\frac 12 tr([\rho^{1/2}, A]^2) \\
&= tr(\rho A^2) -tr (A\rho^{1/2} A\rho^{1/2})
\end{aligned}
\end{eqnarray}
and so the terms we introduced are related to the Wigner-Yanase
information. See Section 3 for the details.

In order to show that the uncertainty relation (\ref{1.3}) and
(\ref{1.4}) are optimal in some special situations, we consider
the position and momentum operators as a pair of conjugate
observables in $L^2(\R)$, and choose the density operator $\rho$
corresponding the thermal state (quasi-free state) for quantum
harmonic oscillator. In this case, we show that the equalities
in(\ref{1.3}) and (\ref{1.4}) hold. See Theorem \ref{theorem4.2}.

Let us describe the main idea employed in this paper. Let $A$ and
$B$ self-adjoint operators (observables) acting on a separable
Hilbert space. Let $\langle \cdot, \cdot \rangle$ be the
Hilbert-Schmidt inner product defined on the class of Hilbert
Schmidt operators :
$$
\langle A, B\rangle := tr(A^* B).
$$
Then the left hand side of (\ref{1.2}) equals to $| \langle A_0
\rho^{1/2} , B_0 \rho^{1/2} \rangle |^2$. In order to make $\rho$
to play same role as $ A$ and $B$, we introduce orthogonal
decompositions
\begin{eqnarray} \label{1.6}
\begin{aligned}
A\rho^{1/2} &= A_{\rho, +} + A_{\rho, -} , \\
B\rho^{1/2} &= B_{\rho, +} + B_{\rho, -}
\end{aligned}
\end{eqnarray}
where
\begin{eqnarray} \label{1.7}
\begin{aligned}
A_{\rho, +} &:= \frac 12 (A\rho^{1/2} + \rho^{1/2} A), \\
A_{\rho, -} &:= \frac 12 (A\rho^{1/2} - \rho^{1/2} A),
\end{aligned}
\end{eqnarray}
and $B_{\rho, +} $ and $B_{\rho, -}$ are defined analogously.
Notice that $\langle A_{\rho, -}, A_{\rho, +} \rangle  =0$ and
$\langle B_{\rho,-}, B_{\rho, +} \rangle =0$. One observes that
\begin{eqnarray*}
\aligned |\langle A_{\rho, +} , B _{\rho, -} \rangle | &= |\langle
A_{\rho, -} , B _{\rho, +} \rangle | \\
&= \frac 14 |\langle [A, B] \rangle_\rho |,
\endaligned
\end{eqnarray*}
The relation (\ref{1.3}) will be followed from the above relation
and the Schwarz inequality. See the proof of Theorem
\ref{theorem2.1} in Section 3. The proof of (\ref{1.4}) is a
little complicate. Let $S$ be the subspace spanned by $B_{\rho, +}
$ and $B_{\rho, -}$ and let $P_S$ be the projection onto $S$.
Denote by $\| \cdot \|_2$ the norm induced by $\langle \cdot,
\cdot \rangle $. Notice that $\| P_S A \rho^{1/2} \|_2 \le \|
A\rho^{1/2} \|_2 $. We will estimate $\| P_S A_\rho \| _2 $ to
prove the relation (\ref{1.4}). See Section 3 for the details.

There has been several proposes to quantify uncertainty by many
authors. A prominent one is the Shanon entrophy\cite{BM, FS}, and
another one is the Fisher information arising in Statistical
inference \cite{Hal,Luo}. Recently Luo and Zhang \cite{LZ} tried
to characterize uncertainty relations by the Wigner-Yanase skew
information \cite{WY}.

The paper is organized as follows: In Section 2, we list our main
results, Theorem \ref{theorem2.1} and Theorem \ref{theorem2.2}. In
Section 3, we produce the proofs of main theorems by introducing
the concept of orthogonal decompositions of $A\rho^{1/2}$ and
$B\rho^{1/2}$. In Section 4, we give a brief discussion on
possible optimal improvement. Then, we give an example of a mixed
state (and a pair of conjugate observables) for which the
equalities in (\ref{1.3}) and (\ref{1.4}) hold.

\section{Improvement of Uncertainty Relations : Main Results}

In this section we first list our main results, Theorem
\ref{theorem2.1} and Theorem \ref{theorem2.2}, and then give some
remarks on the content of the results.

Let $\cH$ be a separable Hilbert space. Denote by $\cL(\cH)$ the
algebra of all bounded linear operators on $\cH$. An operator $T
\in \cL(\cH)$ is called Hilbert-Schmidt if $tr (T^* T) < \infty$,
where $tr(T^* T)$ is the trace of $T^*T$. The class of all
Hilbert-Schmidt operators is denoted by $\cL_2(\cH)$.

We consider a pair of self-adjoint operators $A$ and $B$ acting on
$\cH$. Denote by $D(A)$ (resp. $D(B)$) the domain of $A$ (resp.
$B$). Let $\rho$ be a density matrix (operator) on $\cH$ ; $\rho
\ge 0$ and $tr(\rho) =1$. In order to care of the domain problems
arising from the unboundedness of $A$ and $B$, we assume that the
properties in the following assumption hold:
\begin{assumption} \label{assumption2.1}
Let $A$ and $B$ be self-adjoint operators acting on a separable
Hilbert space $\cH$ and let $\rho \in \cL(\cH)$ be a density
matrix. We assume that the following properties hold:
\begin{enumerate}
\item[(a)] The inclusions $\rho^{1/2} \cH \subset (D(A) \cap
D(B))$, $ A\rho^{1/2} \cH \subset (D(A) \cap D(B))$ and $B
\rho^{1/2} \cH \subset ( D(A) \cap D(B))$ hold.
\item[(b)] The composition maps $A\rho^{1/2}, B \rho^{1/2},
BA\rho^{1/2}, AB\rho^{1/2}, A^2 \rho^{1/2}$ and $ B^2\rho^{1/2}$
define Hilbert-Schmidt operators on $\cH$.
\item[(c)] There is a dense set $D \subset (D(A) \cap D(B))$ such
that the following inclusion hold : $ AD \subset (D(A)\cap D(B)) $
and $BD \subset (D(A)\cap D(B))$.
\item[(d)] The composition maps $\rho^{1/2}A, \rho^{1/2}B,
\rho^{1/2}AB, \rho^{1/2}BA, \rho^{1/2}A^2$ and $\rho^{1/2}B^2$ are
bounded on $D$. The bounded extensions of those operators, denoted
by same symbols, are Hilbert-Schmidt.
\end{enumerate}
\end{assumption}

We now list our main results. For notational simplicity, put
$$\langle T \rangle_\rho := tr (T\rho), \quad \| T\|^2 _\rho =tr
(T^* T\rho)
$$
for any (unbounded) operator, whenever the expressions in the
above are well defined. Let $[A,B]:=AB-BA$ and $\{ A,B\} :=AB+BA$
be the commutator and anti-commutator of $A$ and $B$ respectively.
The following results are improved versions of the uncertainty
relations (\ref{1.1}) and (\ref{1.2}):

\begin{theorem}\label{theorem2.1}
Let $A$ an $B$ be self-adjoint operators acting on a separable
Hilbert space $\cH$ and let $\rho$ be a density matrix. Under
Assumption \ref{assumption2.1}, the relation
\begin{equation}\label{2.1}
\frac 14 | \langle [A,B] \rangle_\rho|^2 + tr (A\rho^{1/2} A
\rho^{1/2} ) tr (B \rho^{1/2} B \rho^{1/2}) \le \| A\|_\rho^2 \| B
\|_\rho^2
\end{equation}
holds.
\end{theorem}
\begin{theorem}\label{theorem2.2}
Let $A, B$ and $\rho$ be the operators as in Theorem
\ref{theorem2.1}. Under Assumption \ref{assumption2.1}, the
relation
\begin{eqnarray}\label{2.2}
\aligned
 \frac 14 | \langle [A,B] \rangle_\rho|^2 + \frac 14 | \langle \{ A, B \} \rangle_\rho|^2 &+  M(A, B; \rho) \\
  \le \|
A_\rho\|^2 \| B \|_\rho^2
\endaligned
\end{eqnarray}
holds, where $M(A,B; \rho) = max \{ M_1(A,B; \rho), M_1(B, A;
\rho) \} $ and $M_1(A,B; \rho)$ is given by
\begin{equation}\label{2.3}
M_1(A,B; \rho) := \frac14 \frac { ( | \langle [A,B]\rangle_\rho |
\, tr (B \rho^{1/2} B \rho^{1/2} ) )^2 }{\| B\| _\rho^4 - (tr(B
\rho^{1/2} B \rho^{1/2})^2 }
\end{equation}
if $tr (B \rho^{1/2} B \rho^{1/2} ) < \| B \|_\rho^2 $, and $M_1
(A,B;\rho) =0$ otherwise.
\end{theorem}

Under Assumption \ref{assumption2.1}, one can check that each term
in the relations (\ref{2.1}) and (\ref{2.2}) is well defined. It
may be possible to weaken Assumption \ref{assumption2.1} to get
the relations (\ref{2.1}) and (\ref{2.2}). Put
$$
A_0 := A- \langle A\rangle_0 , \quad B_0 := B - \langle B
\rangle_\rho.
$$
If one replace $A$ and $B$ by $A_0$ and $B_0$ in the relations
(\ref{2.1}) and (\ref{2.2}), one can see that Theorem
\ref{theorem2.1} and Theorem \ref{theorem2.2} are improvements of
Heisenberg's uncertainty relation (\ref{1.1}) and Schr\"odinger's
uncertainty relation (\ref{1.2}) respectively. Notice that if
$\rho$ is pure, $ tr(A_0 \rho^{1/2} A_0 \rho^{1/2})= tr (B_0
\rho^{1/2} B_0 \rho^{1/2} ) =0$, and so the relations (\ref{2.1})
and (\ref{2.2}) are reduced to the relations (\ref{1.1}) and
(\ref{1.2}) respectively for any pure states.

It may be worth to give discussions on the content of Theorem
\ref{theorem2.1} and Theorem \ref{theorem2.2} in more details.
\begin{remark} \label{remark2.1} \, (a) As mentioned in
Introduction, the functional
$$
\rho \mapsto tr(K\rho^t K^* \rho^{1-t})
$$
is concave for every $0<t<1$ and $K \in \cL (\cH)$ by Lieb's
concavity theorem \cite{Lie}. Thus  the values of the second term
in l.h.s. of (\ref{2.1}) and the third term in l.h.s. of
(\ref{2.2}) measure the  mixtureness of the state. In addition, if
it can be shown that the above functional increases as the
mixtureness of $\rho$ increases in the sense of Uhlmann
\cite{AU,OP}, then obviously the additional terms in (\ref{2.1})
and (\ref{2.2}) increases as the mixtureness of $\rho$ increases.
However we do not know it yet.

(b) The Wigner-Yanase skew information \cite{WY} for any
observable $A$ and a density matrix $\rho$ is defined by
\begin{eqnarray} \label{2.4}
\aligned I(\rho, A) & := \frac 12 tr ([\rho^{1/2} , A] \, [ A,
\rho^{1/2} ] ) \\
& = tr (A^2\rho) - tr (A\rho^{1/2} A \rho^{1/2} ).
\endaligned
\end{eqnarray}
Thus the terms we introduced in Theorem \ref{theorem2.1} and
Theorem \ref{theorem2.2} are related to the above skew
information. Since $0\le I(\rho, A)$, we see that $tr(A\rho^{1/2}
A\rho^{1/2}) \le \|A\|_\rho^2$ and the equality holds if and only
if $[\rho, A]=0$. If $\rho$ commutes with either $A$ or else $B$,
then $\langle [A,B]\rangle_\rho =0$. Thus, if $A$ and $B$ are
conjugate observables, there does not exist such density matrix,
and the strict inequalities $tr (A\rho^{1/2} A\rho^{1/2} ) < \|
A\|^2_\rho$ and $tr (B\rho^{1/2} B \rho ^{1/2}) < \| B \|_\rho^2$
hold for any conjugate abservables $A$ and $B$.
\end{remark}
\begin{remark} \label{remark2.2}
The inequality (\ref{2.2}) is not optimal. In fact, we discarded
complicate non-negative terms in the derivation of (\ref{2.2}). We
will give a discussion on the optimal lower bound of (\ref{2.2}).
see Theorem \ref{theorem4.1} in Section 4.
\end{remark}
\begin{remark}
As an application of the uncertainty relations (\ref{2.1}) and
(\ref{2.2}), we considered the position and momentum operators on
$L^2 (\R)$ as a pair of conjugate observables and the density
matrix $\rho$ corresponding to the thermal state for quantum
harmonic oscillator. We proved that the equalities in the
uncertainty relations in (\ref{2.1}) and (\ref{2.2}) hold in this
case. See Theorem \ref{theorem4.2}.
\end{remark}

\section{Proofs of Theorem \ref{theorem2.1} and Theorem
\ref{theorem2.2}}

We produce the proofs of Theorem \ref{theorem2.1} and Theorem
\ref{theorem2.2} in this Section. Let $A$ and $B$ self-adjoint
operators and $\rho$ a density matrix satisfying the properties in
Assumption \ref{assumption2.1}. For notational brevity, we write
\begin{equation} \label{3.1}
A_\rho := A\rho^{1/2}, \quad B_\rho:=B \rho^{1/2}.
\end{equation}
We decompose $A_\rho$ and $B_\rho$ as
\begin{eqnarray} \label{3.2}
\aligned A_\rho &= A_{\rho,+} + A_{\rho,-}, \\
B_\rho &= B_{\rho,+} + B_{\rho,-},
\endaligned
\end{eqnarray}
where
\begin{eqnarray}\label{3.3}
\aligned A_{\rho,+} &:= \frac 12 (A\rho^{1/2} + \rhoe A) = \frac12
\{ A , \rho^{1/2} \}, \\
A_{\rho,-} &:= \frac 12 (A\rho^{1/2} - \rhoe A) = \frac 12 [A ,
\rho^{1/2} ], \\
B_{\rho,+} &:= \frac 12 (B\rho^{1/2} + \rhoe B) = \frac12 \{ B ,
\rho^{1/2} \}, \\
B_{\rho,-} &:= \frac 12 (B\rho^{1/2} - \rhoe B) = \frac 12 [B ,
\rho^{1/2} ].
\endaligned
\end{eqnarray}
Denote by $\langle T, S \rangle , T, S \in \cL_2(\cH)$, the
Hilbert-Schmidt inner product on $\cL_2(\cH)$ :
\begin{equation} \label{3.4}
\langle T, S \rangle := tr (T^* S), \quad \forall T, S \in
\cL_2(\cH),
\end{equation}
and $\| T\|_2$ the induced norm:
\begin{equation} \label{3.5}
\| T\|_2^2 := tr (T^* T).
\end{equation}
Here we have used the norm $\| T\|_2$ to distinguish it from the
operator norm $\| T\|$.
  In the sequel, we assume that the properties in Assumption
  \ref{assumption2.1} hold.
\begin{lemma} \label{lemma3.1}
  $\quad$(a) $(A\rho^{1/2} )^* = \rho ^{1/2} A$ and $(B\rhoe )^*
  =\rhoe B$.\\[3pt]
 \noindent (b) The equalities
  \begin{eqnarray*}
  \rhoe (A^2 \rhoe) = (\rhoe A) (A\rhoe), && \rhoe (B^2
  \rhoe ) = (\rhoe B) (B\rhoe), \\
  \rhoe (AB \rhoe) = (\rhoe A) (B\rhoe), && \rhoe
  (B A   \rhoe ) = (\rhoe B) (A\rhoe)
  \end{eqnarray*}
hold, where $\rho^{1/2} (A^2 \rho^{1/2})$ is the composite map
(operator product) of $\rho^{1/2}$ and $A^2 \rho^{1/2}$, and
$(\rho^{1/2} A) (A\rho^{1/2})$  the composite map of $\rho^{1/2}
A$ and $ A\rho^{1/2}$, etc.

  \noindent (c) $ \langle  \Arp, \Arm \rangle =0$ and $\langle \Brp, \Brm
  \rangle =0$.
\end{lemma}
\begin{proof}
(a) By Assumption \ref{assumption2.1} (b) -(c), one has that for
any $\varphi \in D$ and $\eta \in \cH$
$$ ( \varphi, A \rhoe \eta)
= (\rhoe A \varphi, \eta).
$$
It follows from Assumption \ref{assumption2.1} (d) that the above
equality holds for any $\varphi, \eta \in \cH$ and so $( A\rhoe
)^* = \rhoe A $. The method used for $A\rhoe$ yields $(B \rhoe)^*
= \rhoe B$.

(b) Those equalities follow from Assumption \ref{assumption2.1}
(b) and the part (a) of the Lemma.

(c) The part (c) of the Lemma follows from the definitions of
$A_{\rho, \pm}$ and $B_{\rho, \pm}$ in (\ref{3.3}) and the trace
property; $tr(TS) =tr (ST)$.
\end{proof}

It follows from Lemma \ref{lemma3.1} (c) that the decompositions
in (\ref{3.2}) are orthogonal decompositions. Also Lemma
\ref{lemma3.1} (a) shows that $\Arp$ and $\Brp$ are self-adjoint,
and $(\Arm)^* = - \Arm$ and $(\Brm)^* =- \Brm$.

\begin{lemma} \label{lemma3.2}
The equalities
\begin{eqnarray*}
\aligned
\langle \Arpm, \Arpm \rangle &= \frac 12 \{ tr(A^2 \rho)
\pm tr (A \rhoe A\rhoe)\} , \\
\langle \Brpm, \Brpm \rangle &= \frac 12 \{ tr(B^2 \rho) \pm tr (B
\rhoe B\rhoe)\} ,\\
\langle \Brpm, \Arpm \rangle &= \frac 14 \{ tr(\{ A, B\} \rho) \pm
2 tr (B \rhoe A\rhoe)\} , \\
\langle \Brpm, A_{\rho,\mp} \rangle &= \frac 14 \{ tr([B,A]\rho)
\endaligned
\end{eqnarray*}
hold.
\end{lemma}
\begin{proof}
The equalities follows from the definitions of $A_\pm$ and $B_\pm$
in (\ref{3.3}) and direct computations.
\end{proof}
Notice that by the first and second equalities in Lemma
\ref{lemma3.2}, one has that
\begin{eqnarray}\label{3.6}
\aligned \| \Arpm \|_2^2 &= \frac 12 \{ \| A\|_\rho^2 \pm tr (A
\rhoe A\rhoe ) \} , \\
 \| \Brpm \|_2^2 &= \frac 12 \{ \| B\|_\rho^2 \pm tr (B
\rhoe B\rhoe ) \}.
\endaligned
\end{eqnarray}
Also one recognizes that the Wigner-Yanase skew information
$I(\rho, A) $ and $\| \Arm\|$ are related by
\begin{equation} \label{3.7}
I(\rho, A) = 2 \| \Arm \|^2.
\end{equation}
See the definition of $I(\rho, A)$ in (\ref{2.4}).

We are now ready to prove Theorem \ref{theorem2.1} and Theorem
\ref{theorem2.2}. It follow from (\ref{3.1}) that
\begin{eqnarray*}
\aligned \langle B_\rho, A_\rho \rangle & = tr (BA \rho) \\
& = \frac 12 tr (\{ B,A \} \rho) + \frac 12 tr ([B,A ] \rho ).
\endaligned
\end{eqnarray*}
Since the first term in the r.h.s. of the above is real and the
second term is pure imaginary,
\begin{equation} \label{3.8}
| \langle B_\rho , A_\rho \rangle |^2 = \frac 14 | tr (\{ B,
A\}\rho )|^2 + \frac 14 | tr ([B, A] \rho )|^2,
\end{equation}
and so by the Schwarz inequality, the Schr\"odinger uncertainty
relation
\begin{equation} \label{3.9}
\frac 14 (tr ([B,A] \rho) |^2 + \frac 14 |tr (\{ B, A \} \rho)|^2
\le \| A_\rho \|_2^2 \| B_\rho\|_2 ^2
\end{equation}
holds. Recall that $\| A_\rho \|_2^2 = \| A\|_\rho^2$ and $\|
B_\rho \|_2^2 = \| B\|_\rho^2$.

\vspace*{0.3cm} \noindent {\it Proof of Theorem \ref{theorem2.1}.}
It follows from the Schwarz inequality and (\ref{3.6}) that
\begin{eqnarray*}
\aligned |\langle \Brp, \Arm \rangle |^2 &\le \| \Brp \|^2_2 \,\|
\Arm \|_2^2 \\
& = \frac 14 (\|B \|_\rho^2 + tr (B\rhoe A\rhoe )) (\| A\|_\rho^2
- tr (A\rhoe A\rhoe )),
\endaligned  \\
\aligned |\langle \Brm, \Arp \rangle |^2 &\le \| \Brm \|^2_2 \,\|
\Arp \|_2^2 \\
& = \frac 14 (\|B \|_\rho^2 - tr (B\rhoe A\rhoe )) (\| A\|_\rho^2
+ tr (A\rhoe A\rhoe )).
\endaligned
\end{eqnarray*}
Thus, by the fourth equality in Lemma \ref{lemma3.2} and the above
inequality. we have
\begin{eqnarray*}
\aligned \frac18 | tr ([B,A]\rho)|^2 &= | \langle \Brp, \Arm
\rangle|^2 + | \langle \Brm, \Arp\rangle |^2 \\
& \le \frac 12 \{ \| B\|_\rho^2 \| A\|_\rho^2 - tr ( B\rhoe B\rhoe
) tr (A\rhoe A\rhoe )\}.
\endaligned
\end{eqnarray*}
The above relation equals to that in Theorem \ref{theorem2.1}.
$\quad \square$.

Now, let us turn to the proof of Theorem \ref{theorem2.2}. Recall
that the class $\cL_2(\cH)$ of all Hilbert-Schmidt operator is a
Hilbert space with the Hilbert-Schmidt inner product $\langle
\cdot, \cdot \rangle $ defined in (\ref{3.4}). Denote by $
\hat{A}_{\rho,\pm}$ and $\hbrpm$ the normalized vectors in
$\cL_2(\cH)$ in the direction of $\Arpm$ and $\Brpm$ respectively
:
\begin{equation} \label{3.10}
\harpm = \Arpm / \| \Arpm \|_2 , \quad \hbrpm = \Brpm / \| \Brpm
\|_2.
\end{equation}
If $\| \Arm \|_2 =0$ (resp. $\| \Brm \|_2 =0)$, we set $\harm = 0$
(resp. $ \hbrm =0$). By (\ref{3.2}) and (\ref{3.10}),
\begin{eqnarray}\label{3.11}
\aligned A_\rho &= \| \Arp \|_2 \harp + \| \Arm \|_2 \harm, \\
B_\rho &= \| \Brp \|_2 \hbrp + \| \Brm \|_2 \hbrm.
\endaligned
\end{eqnarray}
We introduce vectors orthogonal to $A_\rho$ and $B_\rho$ by
\begin{eqnarray}\label{3.12}
\aligned
 A_\rho ^ \perp &= \| \Arm \|_2 \harp - \| \Arp \|_2 \harm, \\
B_\rho ^\perp &= \| \Brm \|_2 \hbrp - \| \Brp \|_2 \hbrm.
\endaligned
\end{eqnarray}
It is easy to check that
\begin{eqnarray} \label{3.13}
\aligned \| A_\rho ^\perp \| _2 = \| A_\rho\|_2 , & \quad \|
B_\rho
^\perp \| _2 = \| B_\rho \|_2, \\
\langle A_\rho^\perp , A_\rho \rangle =0 ,  & \quad  \langle
B_\rho ^\perp , B_\rho \rangle =0 .
\endaligned
\end{eqnarray}
Denote by $\tilde{S}$ and $S$ the subspace of $\cL_2(\cH)$ spanned
by $\{ \harp, \harm \}$ and $\{ \hbrp, \hbrm\}$ respectively, and
let $P_S$ be the projection to $S$.
\begin{proposition} \label{proposition3.1}
The inequality
$$
| \langle B_\rho, A_\rho \rangle |^2 + | \langle  B_\rho ^\perp ,
A_\rho \rangle |^2 \le \| B_\rho \| _2^2 \, \| A_\rho \| _2^2
$$
holds.
\end{proposition}
\begin{proof}
Let $\hat{B}_\rho$ and $\hat{B}_\rho^\perp$ be normalized vectors
in the directions of $B_\rho$ and $B_\rho ^\perp$ respectively :
$$
\hat {B}_\rho = B_\rho / \| B_\rho \|_2, \quad \hat{B} _\rho^\perp
= B_\rho ^\perp / \| B_\rho \|_2 .
$$
Since $\{ \hat{B}_\rho, \hat{B}_\rho^\perp \}$ is an orthonormal
basis of $S$, we have
\begin{equation}
\label{3.14} \| P_S A_\rho \|^2 = | \langle \hat{B}_\rho, P_S
A_\rho \rangle |^2 + | \langle \hat{B}_\rho^\perp , P_S A_\rho
\rangle |^2
\end{equation}
and so
\begin{eqnarray*}
\aligned |\langle \hat{B}_\rho, A_\rho \rangle |^2 + |\langle
\hat{B}_\rho^\perp, A_\rho \rangle |^2 & = \| P_S A_\rho \|^2_2 \\
&\le \| A_\rho \|_2^2.
\endaligned
\end{eqnarray*}
By multiplying $\| B_\rho \|_2^2 $ to the both sides of the above
inequality, we proved the lemma.
\end{proof}

\begin{lemma} \label{lemma3.3}
The equality
$$
|\langle B_\rho^\perp , A \rangle |^2 = M_1 (A,B; \rho ) + M_2 (A,
B ; \rho ),
$$
holds, where $M_1(A,B; \rho)$ is given by (\ref{2.3}) in Theorem
\ref{theorem2.2} and
\begin{eqnarray}\label{3.15}
& & M_2 (A,B;\rho) \\
& &= \frac 14 (\| \Brp \|_2\, \| \Brm \|_2 )^{-2} \big [  \|
B\|_\rho^2 tr (B \rhoe A \rhoe ) -\frac12 \langle \{ B, A\}
\rangle_\rho tr (B\rhoe B \rhoe )  \big]^2 \nonumber
\end{eqnarray}
if $tr (B\rho^{1/2} B \rho^{1/2} ) < \| B\|_\rho^2 $, and $M_2
(A,B ; \rho) =0$ otherwise.
\end{lemma}
\begin{proof}
By the definition of $B_\rho^\perp$ in (\ref{3.12}),
\begin{eqnarray*}
\aligned \langle B_\rho^\perp , A_\rho \rangle & = \langle \| \Brm
\|_2 \hbrp - \|\Brp \|_2 \hbrm , \Arp + \Arm \rangle \\
& = \big\{ \frac {\|\Brm\|_2}{\|\Brp\|_2} \langle \Brp , \Arp
\rangle - \frac {\|\Brp\|_2}{\|\Brm\|_2} \langle \Brm , \Arm
\rangle\big\} \\
& \,\,\,\, + \big\{ \frac {\|\Brm\|_2}{\|\Brp\|_2} \langle \Brp ,
\Arm \rangle - \frac {\|\Brp\|_2}{\|\Brm\|_2} \langle \Brm , \Arp
\rangle\big\}.
\endaligned
\end{eqnarray*}
Since the first term in r.h.s. of the last equality in the above
is real and the second term is pure imaginary, we have
$$
|\langle B_\rho^\perp , A_\rho |^2 = M_1 (A, B ; \rho ) + M_2
(A,B;\rho),
$$
where
\begin{eqnarray}\label{3.16}
\aligned M_1(A, B;\rho) &= \left | \frac {\|\Brm\|_2}{\|\Brp\|_2}
\langle \Brp , \Arm \rangle - \frac {\|\Brp\|_2}{\|\Brm\|_2}
\langle \Brm , \Arp \rangle  \right|^2 , \\
M_2(A, B;\rho) &= \left | \frac {\|\Brm\|_2}{\|\Brp\|_2} \langle
\Brp , \Arp \rangle - \frac {\|\Brp\|_2}{\|\Brm\|_2} \langle \Brm
, \Arm \rangle  \right|^2 .
\endaligned
\end{eqnarray}
By Lemma \ref{lemma3.2},
$$
M_1 (A,B ;\rho ) = \frac 1 {4^2} | \langle [B, A]\rangle_\rho|^2 (
\| \Brp\|_2 \, \| \Brm \|_2 )^{-2} ( \| \Brm \|_2^2 - \| \Brp\|
)^2,.
$$
Substituting (\ref{3.6}) into the above expression, we proved that
$M_1(A,B; \rho)$ in the above equals to that in (\ref{2.3}).

Next, we consider $M_2 (A, B;\rho)$ in (\ref{3.16}).
$M_2(A,B;\rho)$ can be expressed as
$$
M_2(A,B;\rho) = (\|\Brp \|_2 \, \|\Brm\|_2 )^{-2} \big [ \|\Brm
\|_2^2 \langle \Brp , \Brm \rangle - \| \Brp \|_2^2 \langle \Brm,
\Arm \rangle  \big]^{2} .
$$
Using Lemma \ref{lemma3.2} and (\ref{3.6}), one can check that the
above expression equals to that in (\ref{3.15}). Notice that, if
$\|\Brm \|_2 =0$, then $B_\rho^\perp =0$ by (\ref{3.12}). Thus
$M_1(A,B;\rho) =M_2(A,B;\rho)=0$ in this case. This proved the
lemma completely.
\end{proof}

\vspace*{3pt} \noindent {\it Proof of Theorem \ref{theorem2.2}.}
Since $M_2 (A,B ;\rho) \ge 0$, Theorem \ref{theorem2.2} for
$M_1(A,B;\rho)$ follows from Proposition \ref{proposition3.1},
(\ref{3.8}) and Lemma \ref{lemma3.3}. By interchanging the role of
$A_\rho$ and $B_\rho$, we proved the theorem completely. $\quad
\square$.

\section{Optimal Improvement and Application}

We give a brief discussion on the optimal improvement of Theorem
\ref{theorem2.2} which can be obtained by the method used in
Section 3. Then, as an application of Theorem \ref{theorem2.1} and
Theorem \ref{theorem2.2}, we consider  the thermal states of
quantum harmonic oscillator.

\subsection{ Possible Optimal Improvement}

Recall that $S$ is the subspace of $\cL_2(\cH)$ spanned by $\{
\hbrp , \hbrm \}$ and $P_S$ is the projection onto $S$. In the
proof of Theorem \ref{theorem2.2}, we have used the identity
(\ref{3.14}). The quantity $|| P_S A_\rho\|$ is the length of the
projection of $A_\rho$ onto $S$. Thus it is clear that, in order
to obtain the optimal improvement one has to find the vector $X$
with $\| X\|_2 = \| A_\rho \|_2$ in the subspace $\tilde{S} $
spanned by $\{ \harp, \harm \}$, which has the biggest component
in $S$.

In order to find such a vector in $\tilde{ S}$, put
$$
X = \alpha A_\rho +\beta A_\rho^\perp,
$$
where $ \alpha$ and $\beta$ are complex constants satisfying
$$
| \alpha |^2 + |\beta|^2 =1
$$
One has that
\begin{eqnarray*}
\aligned \| P_S X\|_2^2 &= \alpha^2 \| P_S A_\rho \|_2^2 +
\bar{\alpha} \beta \langle P_S A_\rho, P_S A_\rho^\perp \rangle \\
&\,\,+{\alpha} \bar{\beta } \langle P_S A_\rho ^\perp, P_S A_\rho
\rangle + | \beta|^2 \| P_S A_\rho ^\perp \|_2^2.
\endaligned
\end{eqnarray*}
One can choose $\alpha$ such that $\alpha \ge 0$. The first and
last terms in r.h.s. of the above are non-negative. To make the
other terms non-negative, we choose $\beta$ as
$$
\beta =\gamma \langle P_S A_\rho^\perp , P_S A_\rho \rangle / |
\langle P_S A_\rho^\perp, P_S A_\rho \rangle |,
$$
where $\gamma \ge 0$. We then have
\begin{equation} \label{4.1}
\| P_S X\|_2^2 = \alpha^2 \| P_S A_\rho \|_2^2 + 2 \alpha \gamma |
\langle P_S A_\rho, P_S A_\rho ^\perp \rangle | + \gamma^2 \| P_S
A_\rho^\perp \|_2^2,
\end{equation}
where $\alpha$ and $\gamma$ are non-negative real numbers
satisfying
\begin{equation} \label{4.2}
\alpha^2 +\gamma^2 =1.
\end{equation}
Thus the problem is to maximize (\ref{4.1}) under the constrain
(\ref{4.2}). The problem can be solved by the method of the
Lagrange multiplier.

We use the following notation :
\begin{equation} \label{4.3}
a= \| P_S A_\rho \|_2^2 , \,\, b =|| P_S A_\rho^\perp \|_2^2 ,
\,\, c= | \langle P_S A_\rho, P_S A_\rho^\perp \rangle |.
\end{equation}
Put
\begin{equation}\label{4.4}
d= (a-b ) / 2c.
\end{equation}
The method of the Lagrange multiplier implies that
\begin{eqnarray*}
\aligned a \alpha + c \gamma &= \lambda \alpha \\
b \gamma + c \alpha &= \lambda \gamma,
\endaligned
\end{eqnarray*}
where $\lambda$ is the Lagrange multiplier. The above relations
imply
\begin{equation}\label{4.5}
\alpha^2 - 2d \alpha \gamma - \gamma^2 =0.
\end{equation}
Since $\alpha >0$, one has that
\begin{equation} \label{4.6}
\alpha = (d + \sqrt{d^2 +1} ) \gamma.
\end{equation}
From (\ref{4.2}) and (\ref{4.6}), $\alpha$ and $\gamma$ can be
solved explicitly. One may check that
\begin{eqnarray}
\gamma^2& = &1 / (1 + (d + \sqrt{d^2+1 } )^2 ) \nonumber\\
& =& 1/ \big[ 2(d^2 +1) + 2d \sqrt{d^2+1}\, \big]. \label{4.7}
\end{eqnarray}
The relation (\ref{4.5}) and (\ref{4.2}) imply
\begin{equation} \label{4.8}
\alpha \gamma = ( 1- 2\gamma^2 ).
\end{equation}
We substitute (\ref{4.7}) and (\ref{4.8}) into
$$
\| P_S X \|_2^2 = a( 1-\gamma^2 ) + 2c \alpha \gamma + b \gamma^2
$$
to obtain
\begin{eqnarray}
\nonumber \|P_S X ||_2^2 & = & a + c (\sqrt{d^2 +1} -d ) \\
&=& a + \frac 12 \{ [ (a-b)^2 + 4c^2 ] ^{1/2} - (a-b)
\}.\label{4.9}
\end{eqnarray}
We leave that detailed derivation of (\ref{4.9}) to the reader.

Let us denote by
\begin{equation} \label{4.10}
m_3 (A, B; \rho) := \frac12  \{ [ (a-b)^2 + 4c^2 ] ^{1/2} -
(a-b)\},
\end{equation}
where $a,b$ and $c$ are given by (\ref{4.3}). Put
\begin{equation}\label{4.11}
M_3(A, B; d) :=  \| B\| _\rho^2 m _3(A, B; \rho).
\end{equation}
We then  above the following result:
\begin{theorem} \label{theorem4.1}
The relation
\begin{eqnarray*}
\aligned \frac 14 | \langle [A,B] \rangle_\rho |^2 + \frac14 |
\langle \{ A, B\} \rangle _\rho |^2 & + \tilde{M} (A,B; \rho) \\
& \le \|A||_\rho^2 \, \| B\|_\rho^2.
\endaligned
\end{eqnarray*}
holds, where
$$
\tilde{M} (A,B;\rho) = max \{ \sum_{k=1}^3 M_k(A,B; \rho ),
\sum_{k=1}^3 M_k(B,A; \rho) \},
$$
and $M_1(A,B;\rho) $, $ M_2(A,B;\rho)$ and $M_3(A,B; \rho)$ are
given as in (\ref{2.3}), (\ref{3.15}) and (\ref{4.11})
respectively.
\end{theorem}
\begin{proof}
It follows from (\ref{4.9}) that
\begin{eqnarray}
\nonumber \| B\|_\rho^2 \,\| P_S A_\rho \|_2^2 + M_3(A,B;\rho) &=&
\| B\|_\rho^2 \,\| P_S X \|_2^2 \\
& \le & \|B\|_\rho^2 \,\| X\|_2^2 \nonumber \\
&= &\| B\|_\rho^2 \,\|A\|_\rho^2. \label{4.12}
\end{eqnarray}
We recall from the (\ref{3.14}) and Lemma \ref{lemma3.3} that
\begin{equation} \label{4.13}
\| B\|_\rho^2 \, \| P_S A_\rho\|_2^2 = | \langle B_\rho, A_\rho
\rangle |^2 +M_1 (A, B ; \rho) + M_2(A, B ; \rho).
\end{equation}
Thus the theorem follows from (\ref{4.12}),  (\ref{4.13}) and
(\ref{3.8}) together with an interchanging the role of $A$ and
$B$.
\end{proof}

Even if $M_3(A,B ; \rho)$ can be expressed explicitly in terms of
$\|\Arpm\|_2$, $\| \Brpm \|_2$, etc,  the expression is complicate
and so we do not present it here.

\subsection{An Application}
In $L^2(\R)$, the momentum operator $P$ and the position operator
$Q$ are represented by
\begin{equation}
\label{4.14} P= i \frac d {dx} , \quad Q=x.
\end{equation}
It is convenient to introduce the annihilation and creation
operators which are defined as
$$
a= \frac 1{\sqrt{2}} (x + \frac d {dx}), \quad a^* =  \frac
1{\sqrt{2}} (x - \frac d {dx}).
$$
Those operators satisfy the canonical commutation relations
\begin{equation} \label{4.15}
[a, a^*] =1, \quad [a,a] = [a, a^*]=0,
\end{equation}
and $P$ and $Q$ can be written as
\begin{equation}
P= \frac i{\sqrt{2}} (a-a^*) , \quad Q= \frac 1{\sqrt{2}} (a+a^*).
\label{4.16}
\end{equation}
Let $N$ be the number operator defined by
\begin{equation}
\label{4.17} N = a^* a.
\end{equation}
The Hamiltonian for quantum harmonic oscillator is given by
\begin{eqnarray}
H &=& \frac 12 (P^2 + Q^2 ) \nonumber \\
&=& N +\frac12. \label{4.18}
\end{eqnarray}
Let $\Omega$ be the ground state of $H$ and let $\cF_0$ be the
dense subset consisting of finite linear combinations of vectors
$\{ (a^*)^n \Omega, n \in \N\}$. Then $\cF_0$ is a common core for
$a, a^* $ and $N$. For the details, we refer to Section 5.2 of
\cite{BR}.

The density operator $\rho$ corresponding to the thermal state is
given by
\begin{eqnarray}\nonumber
\rho &=& \frac 1Z \exp (-\beta H) \\
&=& \frac 1Z \exp(\beta (N+\frac12)), \label{4.19}
\end{eqnarray}
where $Z = tr (\exp(-\beta H))$ and $\beta >0$ the inverse of the
temperature.

\begin{theorem} \label{theorem4.2}
Let $A=P$ and $B =Q$ and let $\rho$ be given by (\ref{4.19}). Then
the properties in Assumption \ref{assumption2.1} hold (with
$D=\cF_0$). Moreover each side of (\ref{2.1}) and (\ref{2.2})
equals to $\cosh^2 (\beta/2)) / 4 \sinh ^2 (\beta/2)$, and so the
equalities in the uncertainty relations in Theorem
\ref{theorem2.1} and Theorem \ref{theorem2.2} hold.
\end{theorem}
\begin{proof}
Let $a^\# _k, k = 1,2,\cdots, n$, be either $a^*$ or else $a$. It
can be checked that
$$\| \prod_{k=1}^n a_k^\# \varphi \| \le \| (N+n+1 ) ^{n/2}
\varphi\|
$$
for any $\varphi \in \cF_0$ \cite{BR}. Thus $ (\prod_{k=1}^n
a_k^\#) (N+n+1)^{-n/2} $ is bounded operator for each $n$. Thus
the properties (a) and (b) in Assumption \ref{assumption2.1} hold.

Notice that the equalities
$$ a(N+1) = (N+2) a, \,\,\, a^* (N+2) = (N+1) a^*
$$
hold on $\cF_0$. The above equalities imply
$$ (N+2)^{-1} a = a (N+1)^{-1},\,\,\, (N+1)^{-1} a^* = a^* (N+2)
^{-1}.
$$
Using the above relations repeatedly, one can check that the
properties  (c) and (d) in Assumption \ref{assumption2.1} hold. We
leave the details to the reader.

Next, we compute each side of (\ref{2.1}) and (\ref{2.2}). A
direct computation shows that
\begin{equation} \label{4.20}
\langle a^* a\rangle_\rho = e^{-\beta} / (1- e^{-\beta}).
\end{equation}
It follows from (\ref{4.20}) and the canonical commutation
relations (\ref{4.15}) that
\begin{eqnarray}
\nonumber \| Q\|_\rho^2 &=& \frac12 tr ((a+ a^*) (a+ a^*) \rho\\
&=& \frac 12 + e^{-\beta} / (1- e^{-\beta}) \nonumber \\
&=& \frac 12 \cosh(\beta/2) / \sinh(\beta/2). \label{4.21}
\end{eqnarray}
The method used in the above gives
\begin{equation} \label{4.22}
\| P\|_\rho ^2 = \frac 12 \cosh(\beta/2) / \sinh (\beta/2).
\end{equation}
It can be checked that
\begin{equation} \label{4.23}
\rhoe a = e^{\beta/2} a \rhoe, \quad \rhoe a^* = e^{-\beta /2} a^*
\rhoe.
\end{equation}
We use (\ref{4.23}) to obtain
\begin{eqnarray}
\nonumber  tr (Q\rhoe Q \rhoe ) &=& \frac 12 tr((a+a^*)
(e^{\beta/2} a + e^{-\beta/2} a^* ) \rho ) \\
&=& \frac 12 tr (e^{-\beta/2} a a^* + e^{\beta/2} a^* a) \nonumber\\
\nonumber &=& \frac 12 \{ e^{-\beta /2} + (e^{\beta /2}+ e^{-\beta
/2})e^{-\beta } / (1- e^{-\beta }) \} \\
\label{4.24} &=&  1/ 2 \sinh(\beta/2), \\
tr(P\rhoe P\rhoe) &=& 1 / 2 \sinh(\beta/2).\label{4.25}
\end{eqnarray}
A direct computation yields
\begin{equation}  \label{4.26}
\langle \{ P, Q\} \rangle_\rho =0, \quad tr (P \rhoe Q\rhoe) =0.
\end{equation}
Thus (\ref{4.21}) and (\ref{4.22}) imply
\begin{equation} \label{4.27}
\| P\|_\rho^2 \| Q\|_\rho^2 = \cosh^2 (\beta/2) / 4\sinh^2
(\beta/2).
\end{equation}
Since $[P, Q]= i$, (\ref{4.24}) and (\ref{4.25}) imply that
\begin{eqnarray}
\nonumber \text{ l.h.s.  of } \,\, (\ref{2.1}) &=& \frac 14 +
\frac 1 { 4 \sinh^2 (\beta/2)} \\
&=& \cosh^2 (\beta/2) / 4 \sinh^2 (\beta/2). \label{4.28}
\end{eqnarray}
Next, we compute the l.h.s. of (\ref{2.2}). We use (\ref{4.21})
and (\ref{4.24}) to obtain
\begin{eqnarray*}
\aligned \| Q\|_\rho^4 - (tr (Q\rhoe Q\rhoe ))^2 & = \frac 14
\left\{ \frac{\cosh^2 (\beta/2)}{\sinh^2(\beta/2) } - \frac 1
{\sinh^2
(\beta/2)} \right\} \\
&= \frac14,
\endaligned
\end{eqnarray*}
and so
\begin{eqnarray*}
\aligned M_1 (P, Q; \rho) & = (tr (Q\rhoe Q\rhoe ))^2 \\
&= 1/4 \sinh^2 (\beta/2).
\endaligned
\end{eqnarray*}
Thus we conclude that
\begin{equation} \label{4.29}
\text{l.h.s. of } \, (\ref{2.2}) = \cosh^2 (\beta/2) / 4 \sinh^2
(\beta/2).
\end{equation}
Combining (\ref{4.27}) - (\ref{4.29}), we complete the proof of
Theorem \ref{theorem4.2}.
\end{proof}

\vspace{0.2cm} \noindent {\bf Acknowledgements}:  This work was
supported by Korea Research Foundation (KRF-2003-005-C00010),
Korean Ministry of Education.


\begin{thebibliography}{99}
\bibitem[AU]{AU} P. M. Alberti and A. Uhlmann, {\it Stochasticity and
partial order, Doubly stochastic maps and unitary mixing}, VEB
Deutscher verlag (1981).
\bibitem[BM]{BM} I. Bialynicki-Birula and J. Mycielski,
Uncertanity relations for information entropy in wave mechanics,
{\it Comm. Math. Phys.} {\bf 44}, 129-132 (1975).
\bibitem[BR]{BR} O. Bratteli and D. W. Robinson, {\it Operator
Algebras and Quantum Statistical Mechanics} 2, Second Edition,
Springer-Verlag (1996).
\bibitem[FS]{FS} G. B Folland and A. Sitaram, The Uncertainty
principle : A mathematical survey, {\it J. Fourier Anal. Appl. }
{\bf 3}, 207-236 (1997).
\bibitem[Hal]{Hal} M. J. W. Hall, Exact Heisenberg uncertainty
relations, {\it Rhys. Rev. A} {\bf 64}, 052103 (2001).
\bibitem[Hei]{Hei} W. Heisenberg, \"Uber den anschaulichen Inhalt der
quantumtheoretischen Kienematik und Mechanik, {\it Z. Phys. } {\bf
43}, 172-198 (1927).
\bibitem[Lie]{Lie} E. H. Lieb, Convex trace functions and the
Wigner-Yanase-Dyson conjecture, {\it Adv. Math. } {\bf 11},
267-288 (1973).
\bibitem[Luo]{Luo} S.L.Luo, Quantum Fisher information amd
uncertainty relations, {\it Lett. Math. Phys.} {\bf 53}, 243-251
(2000).
\bibitem[LZ]{LZ} S. L. Luo and Z. Zhang, An informational
characterization of Schr\"odinger,s uncertainty relations, {\it J.
Stat. Phys.} {\bf 114}, 1557-1576 (2004).
\bibitem[NC]{NC} N. A. Nielsen and I. L. Chung, {\it Quantum
Computation and Quantum Information}, Combridge University Press
(2002).
\bibitem[OP]{OP} M. Ohya and D. Petz, {\it Quantum entropy and its use},
Springer-Verlag (1993).
\bibitem[Sch]{Sch} E. Schr\"odinger, About Heisenberg uncertainty
relation, {\it Proc. Prussian Acad. Sci. }, Phy-Math. Section XIX,
293-303 (1930).
\bibitem[Suk]{Suk} A. D. Sukhanov, Schr\"odinger uncertainty
relations and physical features of correlated-coherent state, {\it
Theor. Math. Phys.} {\bf 132} , 1277-1294 (2002).
\bibitem[WY]{WY} E. D. Wigner and M. M. Yanase, Information
contents of distribution, {\it Proc. Nat. Acad. Sci. USA}, {\bf
49}, 910-918 (1963).
\end{thebibliography}
\end{document}